\documentstyle[aps,floats,manuscript,epsfig]{revtex}
\begin{document}
\draft
\preprint{\vbox{\noindent
          \null\hfill  INFNCA-TH0205}}
\title{
High-energy neutrino conversion into electron-W pair in magnetic field and
its contribution to neutrino absorption
      }
\author{Andrea Erdas$^{1,2,3,}$\cite{email1} and
        Marcello Lissia$^{2,1,}$\cite{email2}
       }
\address{
$^{1}$ Dipart. di Fisica dell'Universit\`a di Cagliari, S.P. Sestu Km~1,
       I-09042 Monserrato (CA), Italy
\\
$^{2}$ Ist. Naz. Fisica Nucleare (I.N.F.N.) Cagliari, S.P. Sestu Km~1,
       I-09042 Monserrato (CA), Italy
\\
$^{3}$ Department of Physics and Astronomy, The Johns Hopkins University,
       Baltimore, MD 21218
        }
\date{August 9, 2002; revised November 22, 2002}
%
\maketitle
\begin {abstract} 
We calculate the conversion rate of high-energy neutrinos propagating
in constant magnetic field into an electron-$W$ pair ($\nu\to W + e$) from
the imaginary part of the neutrino self-energy. Using the exact propagators
in constant magnetic field, the neutrino self-energy has been calculated
to all orders in the field within the Weinberg-Salam model. We obtain
a compact formula in the limit of $B \ll B_{cr}\equiv m^2/e$.
We find that above the process threshold $E^{(th)} \approx
2.2 \cdot 10^{16}$~eV $\times (B_{cr} / B)$
this contribution to the absorption of neutrinos yields an asymptotic
absorption length $\approx 1.1$~m $\times (B_{cr} / B)^2 \times
(10^{16} \mathrm{eV} / E ) $.
\end {abstract}
\section{Introduction}
The study of creation, propagation, energy loss, and absorption of
neutrinos in magnetic field is important in several astrophysical
contexts and in the early cosmology~\cite{raffeltbook}. Neutrino
self-energy and dispersion relation is modified in magnetized
media~\cite{erdasfeld,erdaskim,erdasisola}, and processes where
neutrinos radiate electron-positron pairs
($\nu\to \nu + e^+ + e^-$)~\cite{borisov,kuznetsov,gvozdev98}
or gammas ($\nu\to \nu + \gamma$)~\cite{gvozdev96,ioannisian,gvozdev98}
have been investigated in the range of energies where it is possible to use
an effective four-fermion interaction and
rates have been obtained in the limits of weak and strong magnetic fields.
As an example of the importance of macroscopic magnetic field as an
effective source of energy loss for energetic neutrinos, we recall that
the
estimate~\cite{borisov} for the rate $\nu \to \nu + e^+ + e^- $
in the strong magnetic field near the surface of a neutron star
is about ten times the rate of pair production in the Coulomb field near the
nucleus of metallic iron.

Conversion of neutrinos in $W$-lepton pairs in the presence of
magnetic fields ($\nu_l\to W + l$) should be considered when
studying the propagation of neutrinos of sufficiently high energies:
we shall show that in this limit this process gives an important
contribution to neutrino absorption. A similar process, $\nu \, \gamma
\to l \, W^{+}$, has been studied by Seckel~\cite{seckel}, who shows that at
energies above the threshold for $W$ production this process is competitive
with
$\nu \, \nu$ scattering at the same center of mass energies.

The process we are considering, where an extremely energetic neutrino
creates a real $W$, is a second order process in the weak coupling
constant $g$, while the radiation of a $e^+ e^-$ pair through a virtual $W$
or a virtual $Z$ is a fourth order process. Therefore, there is an
energy above which the conversion into an electron-$W$ pair
becomes the dominant process.
In addition, since we use the electro-weak lagrangian and not an effective
low-energy theory, our result is valid also at very high
energies, much higher than the
$W$ mass; actually, the $W$-electron decay rate contributes significantly
to neutrino absorption only in this limit.
Notice that eventually the real $W$ decays and that in about
10.5\% of cases the final state contains a
neutrino of the same flavor
({\em e.g.}, $\nu_e\to W + e \to \nu_e  e^+ e^-$), so that the process
can
be thought of as the radiation of a lepton pair; in about
21\% of cases the final state contains
a neutrino of different flavor
({\em e.g.}, $\nu_e \to W + e \to \nu_{\mu}  \mu e$),
and in the remaining 68.5\% of cases the final state contains hadrons.

In  this paper we use Schwinger's proper time method \cite{schwinger} to
calculate the neutrino self-energy in homogeneous magnetic fields and then
we obtain the probability of decay into  $W$-electron pair by extracting
the imaginary part of the self-energy. A similar strategy
was used by Tsai and Erber~\cite{tsai} to extract the photon pair creation
probability from the vacuum polarization in intense homogeneous magnetic
fields.

In Section 2 we briefly review the notation and derive
the one-loop neutrino self-energy in constant magnetic field
\cite{erdasfeld}, in Section 3 we obtain the imaginary part of the
self-energy and the rate of $W$-electron pairs creation in magnetic field.
The ensuing discussion and conclusions are in Section 4.
\section{ Neutrino self-energy in a constant magnetic field}
In this section we review the calculation of the one-loop neutrino
self-energy in a homogeneous magnetic field, using the exact
fermion and  gauge boson propagators in a constant magnetic
field~\cite{erdasfeld}. We consider a magnetic field with magnitude
$B$ pointing along the positive $z$-direction.
The only non-vanishing components of the electromagnetic
field strength tensor $F^{\mu \nu}$ are $F^{12}= -F^{21} = B$. The exact
expression for the electron
$S(x',x'')$ \cite{schwinger,dittrich} and $W$ boson
$G^{\mu \nu}(x',x'')$ \cite{erdasfeld} propagators in a constant
magnetic field are obtained using Schwinger's proper time method:
\begin{equation}
S(x',x'')=\phi^\ast(x',x'')\!\int{d^4k\over (2\pi)^4}e^{ik\cdot (x'-x'')}
S(k) \quad , 
\label{2_01a}
\end{equation}
\begin{equation}
G^{\mu\nu}(x',x'')=\phi(x',x'')\!\int{d^4k\over (2\pi)^4}e^{ik\cdot
(x'-x'')}
G^{\mu\nu}(k) \quad ,
\label{2_02a}
\end{equation}
where the translationally invariant parts of the propagators are
\begin{eqnarray}
\label{2_01}
S(k) =&&  i\int_0^\infty \!\!{ds\over\cos eBs} \\ &\times&
{\exp{\left[-is\left(m^2-i\epsilon+k^2_{\parallel}+
k^2_{\perp}{\tan eBs\over eBs}\right)\right]}}
\left[(m-
\not\! k_{\parallel})e^{-ieBs\sigma_3}-
{\not\! k_{\perp}\over \cos eBs}
\right] \quad , \nonumber
\end{eqnarray}
and
\begin{eqnarray}
G^{\mu \nu}(k)&=&i
\int_0^\infty \!\!{ds\over\cos eBs}\,
{\exp{\left[-is\left(k^2_{\parallel}+
k^2_{\perp}{\tan eBs\over eBs}\right)\right]}}
\Biggl\{
e^{-is(M^2-i\epsilon)}[g^{\mu \nu}_{\parallel}
+(e^{2eFs})^{\mu\nu}_{\perp}]
\nonumber \\
&&+
\left[\left(k^{\mu}+k_{\lambda}
F^{\mu \lambda}{\tan eBs\over B}\right)
\left(k^{\nu}+k_{\rho}
F^{\rho \nu}{\tan eBs\over B}\right)
+i{e\over 2}\left(F^{\mu\nu}-
g^{\mu\nu}_{\perp} B\tan eBs
\right)\right]
\nonumber \\
&&\times
\left({e^{-is(M^2-i\epsilon)}-e^{-is(x M^2-i\epsilon)}\over M^2}\right)
\Biggr\} \quad .
\label{2_02}
\end{eqnarray}
In the rest of the paper we shall drop the infinitesimal imaginary
contribution to the masses $-i\epsilon$, which determines the correct
boundary conditions; if necessary it can be easily reintroduced:
$m^2 \to m^2 -i\epsilon$ and $M^2 \to M^2 -i\epsilon$.
In our notation,
$-e$ and $m$ are the charge  and mass of the electron,
$M$ the $W$-mass, $x$ the gauge parameter, $\sigma_3=\sigma^{12}=
{i \over 2}[\gamma^1, \gamma^2]$, and the metric is
$g^{\mu \nu} = \mathrm{diag}(-1,+1,+1,+1)$.
We choose the electromagnetic vector potential to be
$A_\mu=-{1\over2}F_{\mu \nu}x^\nu$ and, therefore,
the phase factor in Eqs. (\ref{2_01a}) and (\ref{2_02a}),
which is independent of the integration path, is
\cite{dittrich}
\begin{equation}
\phi(x',x'')=\exp\left[
ie\int^{x'}_{x''}dx_\mu A^\mu(x)
\right]=\exp\left(
i {e\over 2}x''_\mu F^{\mu \nu} x'_\nu
\right)  \quad .
\label{2_07}
\end{equation}

For any 4-vector $a^\mu$ we use the notation
$a^\mu_{\parallel}=(a^0,0,0,a^3)$ and
$a^\mu_{\perp}=(0,a^1,a^2,0)$;
it is easy to show that
\begin{equation}
\left(e^{2eFs}\right)^{{\mu}\nu}=
(g^{\mu\nu})_{\perp} \cos{(2eBs)}
+ {F^{\mu\nu} \over B} \sin{2(eBs)} \quad .
\end{equation}
Note that the $W$ and Goldstone scalar propagator were obtained in Ref.
\cite{erdasfeld} by introducing a new gauge fixing term (EGF) which is
manifestly invariant under electromagnetic
gauge transformations. The advantage of the EGF gauge is that
the electromagnetic potential has no cross couplings with the $W$ and
Goldstone fields and, therefore, these two
fields do not mix in the presence of a magnetic field.

In the remainder of this paper we focus our attention on electron-type
neutrinos, the generalization to $\mu$ and $\tau$-neutrino is
straightforward.
For the purpose of this work, it would seem convenient to work in the
unitary gauge ($x \rightarrow \infty$), where the unphysical scalars
disappear. 
However, the $W$-propagator is quite cumbersome in this gauge. We prefer
to work in the Feynman gauge ($x=1$), where the expression of the
propagator is much simpler.
In principle the choice of the Feynman gauge carries the price of
calculating an additional bubble diagram: the one with the Goldstone scalar.
But this scalar bubble diagram does not contribute to leading order, since
it is suppressed by a factor of  $m^2/M^2 \approx 4.04\cdot 10^{-11}$, and
can be neglected.
Therefore, we only need to calculate the bubble diagram with a $W$-boson:
\begin{equation}
{\Sigma}_{W}(x',x'')={ig^2\over 2}\gamma_R{\gamma}_{\mu}
S(x',x''){\gamma}_{\nu}\gamma_L G^{\mu \nu}
(x',x'')
\end{equation}
where
\begin{equation}
\gamma_R={1+\gamma_5\over2} \quad \quad
\gamma_L={1-\gamma_5\over2} \quad .
\end{equation}
This expression is translationally invariant, since the $W$ and the electron
carry opposite charge and therefore
the phase factor $\phi$ in the $W$-propagator cancels the phase factor
$\phi^\ast$ in the electron propagator.
We can write ${\Sigma}_{W}(p)$ in momentum space using
\begin{equation}
{\Sigma}_{W}(x',x'')={\int}\!{{d^4\!p}\over{{(2{\pi})}^4}}
e^{ip{\cdot}(x'-x'')}{\Sigma}_{W}(p)
\end{equation}
as
\begin{eqnarray}
\Sigma_{W}(p)=&&-{ig^2\over 2}
\int\!{{d^4 k}\over{{(2{\pi})}^4}}{\int}^{\infty}_{0}
{ds_1\over{\cos z_1}}{\int}^{\infty}_{0}
{ds_2\over{\cos z_2}}{e^{-is_1(m^2 +q^2_{\parallel}
+q^2_{\perp}{{\tan {z_1}}\over {z_1}})}}
{e^{-is_2(M^2 +k^2_{\parallel}+k^2_{\perp}{{\tan {z_2}}\over
{z_2}})}}{\times}
\nonumber \\
&&\gamma_R{\gamma}_{\mu}
\left[(m-
\not\! q_{\parallel})e^{-iz_1\sigma_3}-
{\not\! q_{\perp}\over \cos z_1}
\right]
[g^{\mu \nu}_{\parallel}
+(e^{2eFs_2})^{\mu\nu}_{\perp}]{\gamma}_{\nu}\gamma_L
\end{eqnarray}
where
\begin{equation}
q = p-k \quad , \quad\quad z_1 = eBs_1 \quad , \quad\quad z_2 = eBs_2 \quad
.
\end{equation}

We now do the straightforward $\gamma$-algebra, change variables from $s_i$
to
$z_i$, translate the $k$ variables
of integration as follows
\begin{equation}
(k_{\parallel}\,\,,\,\,k_{\perp})
\,\,{\rightarrow}\,\,
(k_{\parallel}+{z_1\over z_1+z_2}
p_{\parallel}\,\,,\,\,
k_{\perp}
+ {{\tan{z_1}}\over{{\tan{z_1}}+{\tan{z_2}}}}  p_{\perp})
\end{equation}
and, finally, perform the four gaussian integrals over the shifted variables
$k$. The result is:
\begin{eqnarray}
\Sigma_{W}(p)=&& {g^2\over (4\pi)^2}
{\int}^{\infty}_{0}
{\int}^{\infty}_{0}
{dz_1 dz_2 \over (z_1+z_2) \sin(z_1+z_2)}
e^{-i(m^2/eB)[z_1 + z_2 M^2 / m^2 + (z_1 + z_2)\varphi_0]}  \times
\nonumber \\
&& \left[{z_2\over z_1+z_2}{\not\! p_{\parallel}}e^{
i\sigma_3(z_1+2 z_2)}+{\sin z_2 \over \sin(z_1+z_2)}
{\not\! p_{\perp}}\right]
\!\gamma_L + (\mathrm{c.t.})
\end{eqnarray}
where
\begin{equation}
\varphi_0={z_1 z_2 \over (z_1 + z_2)^2}
\frac{p_{\parallel}^2}{m^2}
+{\sin{z_1} \sin{z_2} \over (z_1+z_2) \sin{(z_1+z_2)}}
\frac{p_{\perp}^2}{m^2} \quad .
\end{equation}
and the appropriate counter-terms (c.t.) are defined such that
\begin{equation}
({\rm c.t.})=-{\Sigma}_W(p){\Bigr|}_{B=0,{\not\!\,p}=0}-{\not\! p}
{\biggl[}{{{\partial}{\Sigma}_W(p)}\over{{\partial}{\not\! p}}}
{\biggr]}_{B=0,{\not\!\, p}=0}.
\end{equation}
The $B$-independent counter-terms are unimportant for the purpose of
this work, since they do not contribute to the imaginary part of the
self-energy.

In order to more easily isolate the range of variables that give most of
the contribution to the integrals and the terms in the integrand that are
negligible, we also find it convenient to change integration
variables from $(z_1, z_2)$ to $(z, u)$
\begin{equation}
z = \frac{m^2}{eB}(z_1 +  z_2) \equiv \frac{1}{\beta}
      (z_1 +  z_2) \quad\quad {\mathrm{and}} \quad\quad
u = \frac{M^2}{m^2} \frac{z_2}{z_1+z_2} \equiv \frac{1}{\eta}
\frac{z_2}{z_1+z_2} \quad ,
\end{equation}
where we have introduced the two parameters $\beta = eB/m^2 $ and
$\eta = m^2/M^2 =4.0376 \cdot 10^{-11}$; the resulting expression is:
\begin{eqnarray}
\label{sigma17}
\Sigma_{W}(p) = &&
 {g^2\over (4\pi)^2} \left(\frac{eB}{M^2}\right)
{\int}^{\infty}_{0}\frac{dz}{\sin{\beta z}}
{\int}^{M^2/m^2}_{0} \!\!\!\! du \,
e^{-i z [ 1 -\eta u + u + \varphi_0]} \times \nonumber \\
&& \left[ \eta u {\not\! p_{\parallel}}e^{
i\sigma_3 \beta z (1+\eta u)}+{\sin{[\beta z \eta u]} \over \sin{\beta z} }
{\not\! p_{\perp}}\right]
\!\gamma_L + (\mathrm{c.t.})
\end{eqnarray}
with
\begin{equation}
\varphi_0= u ( 1- \eta u)
\frac{p_{\parallel}^2}{M^2}
+{\sin{[\beta z (1-\eta u)]} \sin{[\beta z \eta u]}
 \over \eta\beta z \sin{\beta z}}
\frac{p_{\perp}^2}{M^2}.
\end{equation}

\section{ Rate of $W$-electron pair creation}
Because of the oscillating phase $\exp{(-iz)}$, the main contribution
to the integral over $z$ comes from the region where $z \lesssim 1$. If
we are only interested in neutrinos that travel through a ``moderate''
magnetic
field $eB \ll m^2 = eB_{cr}$, this means $\beta \ll 1$, we can expand
the terms in the integrand in power series of $\beta z \ll 1$.
Furthermore, since ${\Sigma}_W$ is quite small, of order $ g^2 eB/M^2$,
as it can be inferred from the expression (\ref{sigma17}) or from the
explicit calculation in Ref.~\cite{erdasfeld}, we can set
${\not\! p}=0$ in ${\Sigma}_W$.

After this is all done, we obtain

\begin{eqnarray}
\label{sigma_31}
\Sigma_{W}(p) \simeq && {g^2\over (4\pi)^2} \left(\frac{eB}{M^2}\right)^2
{\not\! p_{\perp}} \gamma_L
{\int}^{\infty}_{0}dz z
{\int}^{M^2/m^2}_{0} \!\!\! du u
\left[\frac{2}{3} + \eta u + \frac{1}{3} (\eta u)^2 \right] \nonumber \\
&&\times 
e^{-i z[1+u-\eta u + \frac{1}{3}\frac{p^2_{\perp}}{m^2}
(\frac{eB}{M^2})^2 z^2 u^2 (1-\eta u)^2 ]} \quad .
\end{eqnarray}

Notice that the term in the exponential proportional to $z^3$ cannot be
dropped in general, in spite of the fact that $\beta z \ll 1$:
the coefficient of $(\beta z)^2$, {\em i.e.} $(m p_{\perp}/ M^2)^2$,
could be very large when the neutrino is very energetic. In fact, there
are two physically interesting regimes that are discriminated by the
dimensionless field dynamical parameter
\begin{equation}
\label{xi}
\xi \equiv  \frac{e B p_{\perp}}{m M^2} \quad ,
\end{equation}
which can be read from the ratio of the $z$ and $z^3$ term in the
exponential
of Eq.~(\ref{sigma_31}) or inferred from kinematical considerations and the
momentum change of a virtual $e-W$ pair in a magnetic field.

At low energies, $\xi \ll 1$, the $z^3$ term in the exponential can be
dropped, the self-energy is real and we obtain the result of
Ref.~\cite{erdasfeld}; at high energies, $\xi \gg 1$, the self-energy
acquires
an imaginary part that we shall calculate in the following. In this last
case
we write $\Sigma_{W}$ as

\begin{equation}
\label{sigma_33}
\Sigma_{W}(p) = \frac{2}{3}
{g^2\over (4\pi)^2} \left(\frac{eB}{M^2}\right)^2
{\not\! p_{\perp}} \gamma_L
{\int}^{\infty}_{0}dz z
{\int}^{\infty}_{0} \!\!\! du u
e^{-i z[1+ u  + \frac{1}{3} \xi^2  z^2 u^2  ]}
\end{equation}
where we have
dropped all nonleading terms in $\eta u$ and extended the integration in
$d u$ to $\infty$, due to the fact that the main contribution to the
integral in 
$d u$ comes from the region $u \lesssim 1$ because of the oscillating phase
$\exp{(-i z u)}$ and that $\eta = m^2/M^2$ is extremely small.

The integral in $d z$ of the imaginary part of the self-energy can be
performed with the substitution $z= y \sqrt{1+u} / (\xi u)$
in terms of the modified Bessel function
\begin{equation}
K_{2/3}(w) = \sqrt{3}
{\int}^{\infty}_{0}y\sin \left[{3\over 2}w(y+{1\over 3}y^3)\right]dy
\end{equation}
obtaining
\begin{equation}
\label{sigma_35}
\Im \Sigma_{W}(p) = -\frac{2}{3}
{g^2\over (4\pi)^2} \left(\frac{eB}{M^2}\right)^2
{\not\! p_{\perp}} \gamma_L
\frac{1}{\sqrt{3} \xi^2}
{\int}^{\infty}_{0} \!\!\! du \frac{1+u}{u}
K_{2/3}\left[\frac{\sqrt{3}}{\xi}\frac{2}{u}
         \left(\frac{1+u}{3}\right)^{3/2} \right]        \quad .
\end{equation}
The final integration in $du$ yields
\begin{equation}
\label{sigma_36}
\Im \Sigma_{W}(p) = - \frac{g^2}{24\pi} \left(\frac{eB}{M^2}\right)^2
\left(1 + \sqrt{3} \frac{m M^2}{eB p_{\perp}}\right)
e^{-\sqrt{3} \frac{m M^2}{eB p_{\perp}}}  \quad ,
\end{equation}
and, therefore, the absorption coefficient,
$\alpha = -2 p_{\perp} \Im \Sigma_{W}(p)$, is
\begin{equation}
\label{alpha_finale}
\alpha = \frac{g^2}{12\pi \hbar c} \frac{p_{\perp}c}{\hbar c}
\left(\frac{m}{M}\right)^4 \left(\frac{B}{B_{cr}}\right)^2
\left(1 + \sqrt{3} \frac{M}{m} \frac{Mc}{p_{\perp}} \frac{B_{cr}}{B}\right)
e^{-\sqrt{3} \frac{M}{m} \frac{Mc}{p_{\perp}} \frac{B_{cr}}{B}} \quad .
\end{equation}

From the exponential in Eq.~(\ref{alpha_finale}) we can read the threshold
for
the process~\footnote{The kinematical threshold is obviously given by the
sum
of the masses $M+m$, below which the rate is rigorously zero: this threshold
is lost in our expansion, but it is unimportant as long as the effective
threshold, below which the process is exponentially suppressed, is very
much larger than $M+m$.}: if we define $\xi^{(th)} = \sqrt{3}$ then
\begin{equation}
\label{threshold}
cp_{\perp}^{(th)} = \sqrt{3} Mc^2 \frac{M}{m} \frac{B_{cr}}{B} \approx
               2.2 \cdot 10^{16} \mathrm{ eV  }
              \left(\frac{B_{cr}}{B}\right) \quad .
\end{equation}
For energies well above the threshold, the absorption coefficient has
the asymptotic behavior
\begin{equation}
\label{asymp}
\alpha 
=\frac{g^2}{12\pi \hbar c}
    \frac{p_{\perp}c}{\hbar c} \left(\frac{eB}{M^2}\right)^2
= \frac{g^2}{12\pi \hbar c} \frac{p_{\perp}c}{\hbar c}
\left(\frac{m}{M}\right)^4 \left(\frac{B}{B_{cr}}\right)^2
=  0.935
  \left(\frac{p_{\perp}c}{10^{16} \mathrm{eV}}\right)
   \left(\frac{B}{B_{cr}}\right)^2 \mathrm{m}^{-1} \quad ,
\end{equation}
where we have used the numerical value
$G_F /(\hbar c)^3 = g^2/[\sqrt{2}(2Mc^2)^2 \hbar c]=1.16639\cdot 10^{-5}
\mathrm{GeV}^{-2}$.

In case we consider $\nu_{\mu}$ ($\nu_{\tau}$) instead of $\nu_{e}$
the corresponding threshold energies are higher by a factor
$m_{\mu} / m_{e} \approx 206.8$ ($m_{\tau} / m_{e} \approx 3478$),
while the asymptotic behavior does not change, since it depends only
on $M$ and not on $m$~\footnote{In principle the full result would have
additional dependences on the lepton mass coming from terms that contain
$\eta =(m/M)^2$ and that we have disregarded being extremely small for
electrons; these corrections are larger for muons and especially taus,
but still small.}.

\section{Discussion and conclusions}
\begin{figure}[fg1]
\psfig{figure=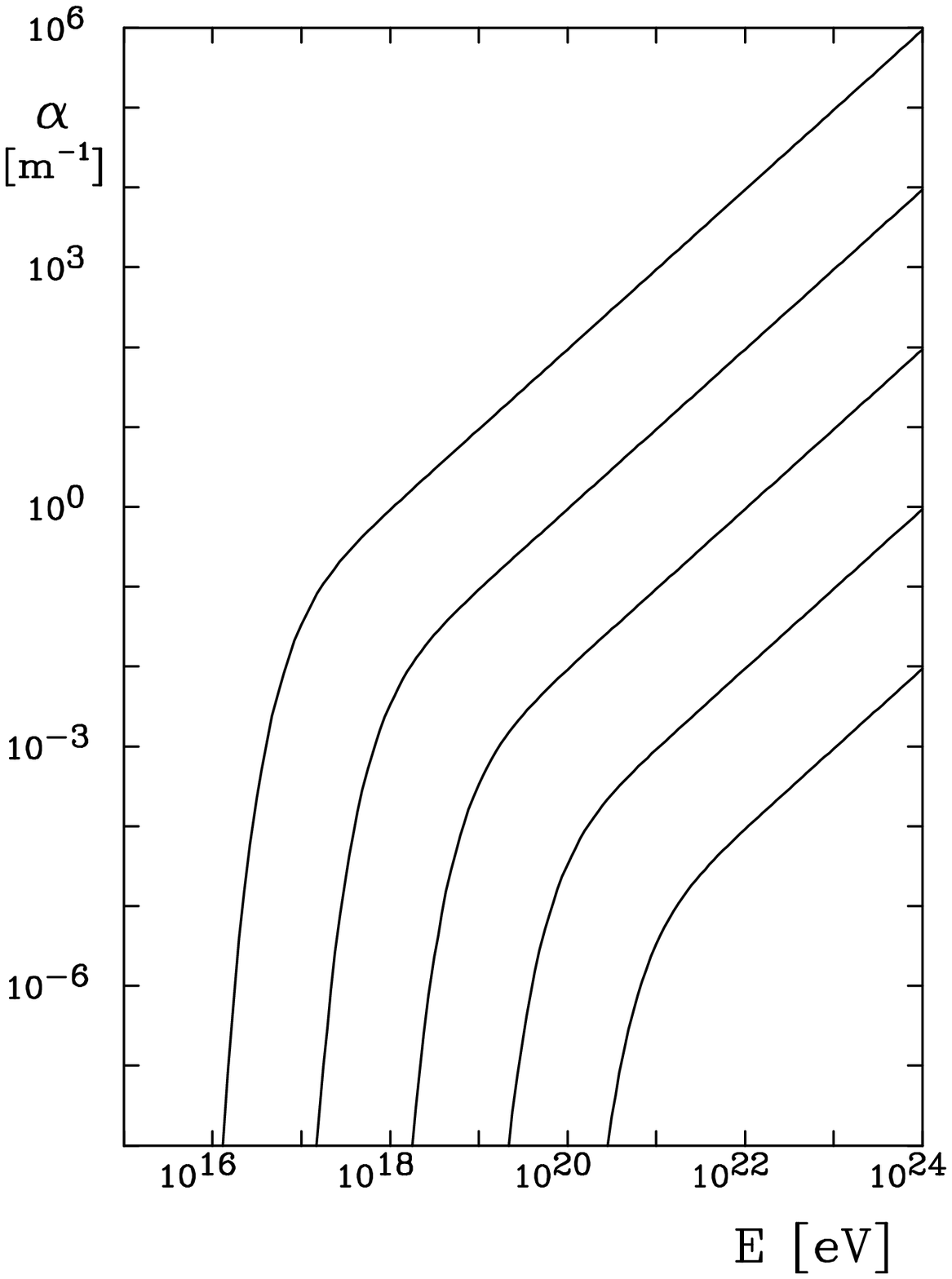,bbllx=100pt,bblly=190pt,bburx=510pt,bbury=740pt,%
height=18cm}
\caption[fc1]{
Neutrino absorption coefficient for the process $\nu_{e}\to W + e$ in
inverse meters as function of the neutrino transverse energy
($E\equiv p_{\perp} c$) in eV for five values of the magnetic field:
$B = 10^{-1}B_{cr}$, $B = 10^{-2}B_{cr}$, $B = 10^{-3}B_{cr}$,
$B = 10^{-4}B_{cr}$, $B = 10^{-5}B_{cr}$ going from top to bottom,
with $B_{cr} = 4.4 \cdot 10^{10}$~Gauss.
The same curves apply for $\nu_{\mu}$ ($\nu_{\tau}$), if both the horizonal
and the vertical scale are multiplied
times $(m_{\mu}/m_{e}) = 206.768266$ ($(m_{\tau}/m_{e})=3477.6 $).
\label{fig1}
           }
\end{figure}

In Figure~\ref{fig1} we show the absorption coefficient
for $\nu_{e}\to W + e$ as function of the
neutrino transverse energy $E\equiv p_{\perp}c$ for several values of
the magnetic field $B$: $B = 10^{-1} B_{cr}$, $B = 10^{-2} B_{cr}$,
$B =  10^{-3} B_{cr}$, $B = 10^{-4} B_{cr}$, and $ B = 10^{-5} B_{cr}$
with $B_{cr} = m^2/e = 4.4 \cdot 10^{10}$~Gauss.
It is evident that the process has an energy threshold that grows for
smaller values of $B$, as quantitatively described by
Eq.~(\ref{threshold}). If we consider the angular coefficient, which is
one, and the spacing between the curves above the threshold, we find that
$\alpha$ grows linearly with $E$ and quadratically with $B$, in agreement
with Eq.~(\ref{asymp}). The same Figure~\ref{fig1} is valid for $\nu_{\mu}$
($\nu_{\tau}$), if one multiplies both the horizonal and the vertical scale
times $(m_{\mu}/m_{e}) = 206.768266$ ($(m_{\tau}/m_{e})=3477.6 $).

\begin{figure}[fg2]
\psfig{figure=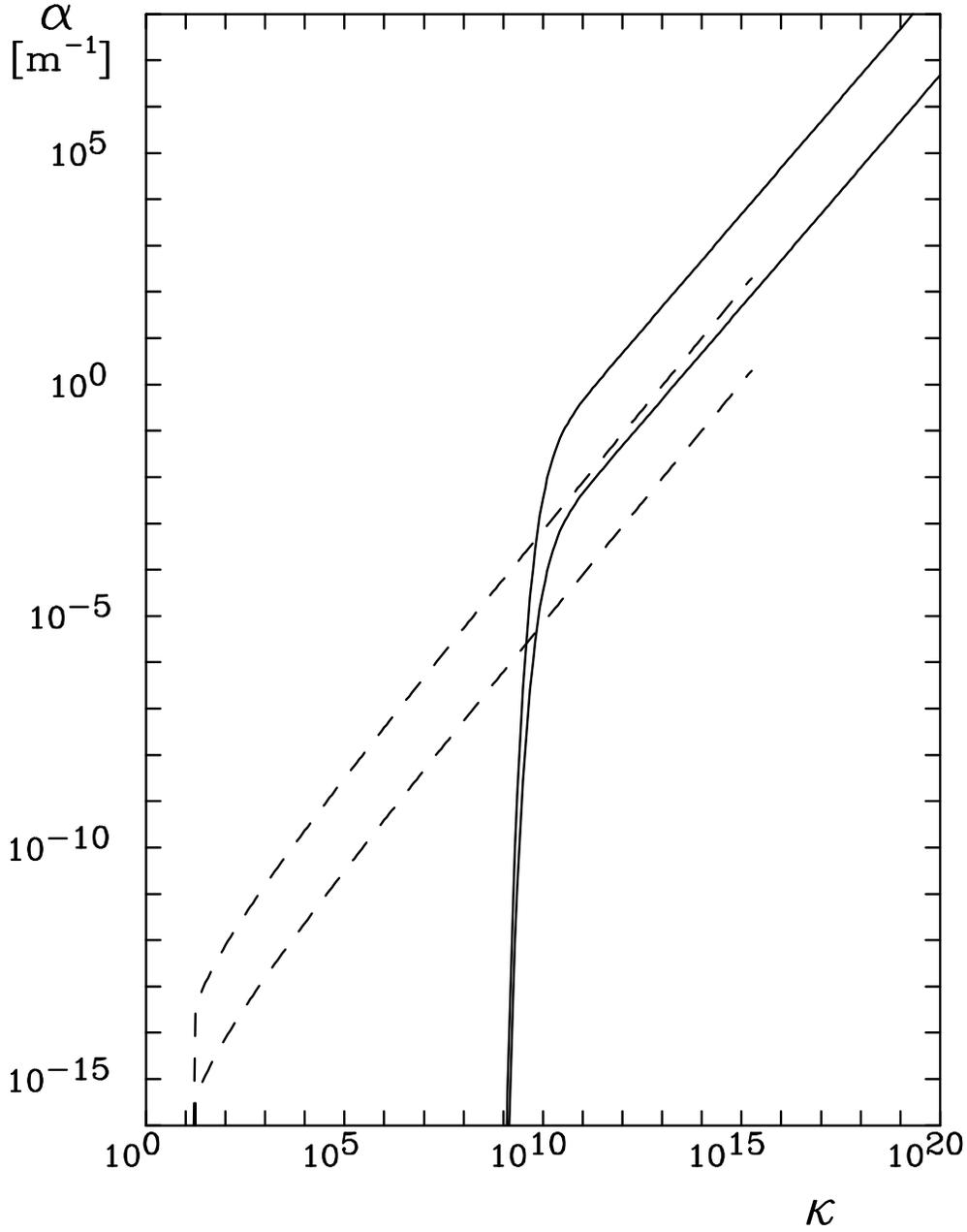,bbllx=100pt,bblly=190pt,bburx=510pt,bbury=740pt,%
height=18cm}
\caption[fc2]{
Neutrino absorption coefficients for the processes $\nu_{e}\to W + e$,
(solid curves) and $\nu\to \nu + e^+ + e^-$ (dashed curves) in inverse
meters as function of the dimensionless characteristic parameter
$\kappa\equiv e B p_{\perp} / m^3$ for two values of the magnetic
field: $B = 10^{-1}B_{cr}$ and $B = 10^{-3}B_{cr}$ going from top to bottom.
\label{fig2}
           }
\end{figure}

In Figure~\ref{fig2} we compare the process $\nu_{e}\to W + e$ with
the process $\nu\to \nu + e^+ + e^-$ that has a much lower threshold;
this threshold can be estimated from the dimensionless field dynamical
parameter
characteristic to this process
$\kappa\equiv e B p_{\perp} / m^3$, which is analogous to $\xi$
of Eq.~(\ref{xi}) with the substitution
$M\to m$: the threshold is smaller by a factor of
about $m^2/M^2\approx 4\cdot 10^{-11}$.
The curves plotted as function of $\kappa$ show clearly that the thresholds
of both processes are function only of the product
$B p_{\perp}$, see Eq.~(\ref{threshold}).
For the process $\nu\to \nu + e^+ + e^-$ we use the result shown in
Eq.~(8) of Ref.~\cite{kuznetsov}:
\begin{equation}
\label{epmkuz}
\alpha(\nu\to \nu e^+ e^-) = \frac{G_F^2 (g^2_V + g^2_A)}{(3\pi)^3} m^4 E
                           \left(\frac{B}{B_{cr}}\right)^2
              \left( \ln(\kappa) - \frac{\ln(3)}{2} -\gamma_E -\frac{29}{24}
              \right)
\quad ,
\end{equation}
which the authors claim valid for $E\ll M^3 / eB$
($\kappa \ll (M/m)^3\approx 4 \cdot 10^{15}$):
we plot Eq.~(8) of Ref.~\cite{kuznetsov}, see Eq.~(\ref{epmkuz}),
up to $\kappa = 10^{15}$.
We show results only for $B= 0.1 B_{cr}$ (top curves) and $B= 0.001 B_{cr}$
(bottom curves). 
We see that the process $\nu\to \nu + e^+ + e^-$ (dashed curves) dominates
below the threshold of $\nu\to W + e $ (solid curves), but above this
threshold $\nu\to W + e $ is almost two orders of magnitude larger (about
a factor 50). Therefore, above the threshold $E \approx (m M^2) / eB$
($\kappa \approx (M/m)^2 \approx  2 \cdot 10^{10}$),
the effective lagrangian (four fermion interaction) cannot be used. The
rate of $\nu\to \nu + e^+ + e^-$ can instead be estimated using the rate
of $\nu\to  W + e$, which gives the total adsorption, times the branching
ratio of $W \to \nu_{e} + e $, which is $(10.66\pm 0.20) \% $.

In conclusion,
we have calculated the absorption rate of very high energy neutrinos in
magnetic field. Our main result is
given by the compact formula in Eq.~(\ref{alpha_finale}) valid for
electron neutrinos. The result for muon or tau neutrinos can be obtained
by substituting the electron mass $m$ with the muon or tau mass remembering
that the electron mass $m$ appears also in $B_{cr}=m^2/e$.

This process is exponentially suppressed, and, therefore, it can be
disregarded, for energies below a threshold energy inversely proportional
to the magnetic field; for a field one tenth of the critical field
this energy is of the order of $10^{17}$~eV, see Eq.~(\ref{threshold}) and
Fig.~\ref{fig1}.

Above this threshold the absorption coefficient grows linearly with energy
and quadratically with the field as shown in Eq.~(\ref{asymp}):
for a field one tenth of the critical field and an energy of $10^{18}$~eV
the absorption coefficient is about one inverse meter (see Fig.~\ref{fig1}).

Above the threshold this process substitutes the radiation of $e^+ e^-$
pairs
as dominant mechanism for $\nu$ absorption in magnetic field
(see Fig.~\ref{fig2}).

\acknowledgements
A. Erdas wishes to thank Gordon Feldman for helpful discussions
and the High Energy Theory Group of the Johns Hopkins University for the
hospitality extended to him during his several visits.

This work is partially supported by M.I.U.R. (Ministero dell'Istruzione,
dell'Universit\`a e della Ricerca) under Cofinanziamento P.R.I.N. 2001.



\end{document}